\documentclass[conference]{IEEEtran}
\IEEEoverridecommandlockouts
\usepackage{tablefootnote}
\usepackage{cite}
\usepackage{amsmath,amssymb,amsfonts}
\usepackage{algorithmic}
\usepackage{graphicx}
\usepackage{textcomp}
\usepackage{xcolor}
\def\BibTeX{{\rm B\kern-.05em{\sc i\kern-.025em b}\kern-.08em
    T\kern-.1667em\lower.7ex\hbox{E}\kern-.125emX}}

\usepackage[linesnumbered,ruled,vlined]{algorithm2e}
\usepackage{epstopdf}
\usepackage{epsfig} 
\usepackage{times} 
\usepackage{amsmath, amssymb, amsfonts} 
\usepackage{float}
\usepackage{booktabs}
\usepackage{multirow}
\usepackage{utfsym} %
\usepackage[utf8]{inputenc}

\usepackage{hyperref}
\hypersetup{
  colorlinks=true,         
  linkcolor=blue, 
  citecolor=black,
  filecolor=magenta,
  urlcolor=black  
}
\usepackage{bbm}
\usepackage{etoolbox}
\makeatletter
\patchcmd{\@IEEEpubidpullup}{\vskip\z@}{\vskip 1cm}{}{}
\newcommand{\linebreakand}{%
\end{@IEEEauthorhalign}
\hfill\mbox{}\par
\mbox{}\hfill\begin{@IEEEauthorhalign}
}
\makeatother
\makeatletter

\begin{document}

\title{
Personalized Continual EEG Decoding: Retaining and Transferring Knowledge
\\

\thanks{This work was partly supported by the Institute of Information \& Communications Technology Planning \& Evaluation (IITP) grant funded by the Korea government (MSIT) (No. RS-2019-II190079, Artificial Intelligence Graduate School Program, Korea University) and the National Research Foundation of Korea (NRF) grant funded by the MSIT (No.2022-2-00975, MetaSkin: Developing Next-generation Neurohaptic Interface Technology that enables Communication and Control in Metaverse by Skin Touch).}
}

\IEEEpubid{%
    \makebox[\columnwidth]{\fontsize{10}{12}\selectfont\fontfamily{ptm}\selectfont%
    979-8-3315-2192-9/25/\$31.00~\copyright~2025 IEEE\hfill}%
    \hspace{\columnsep}\makebox[\columnwidth]{}
}


\author{
\IEEEauthorblockN{~~~~~~~~~Dan Li}
\IEEEauthorblockA{\textit{~~~~~~~~~Dept. of Artificial Intelligence} \\
\textit{~~~~~~~~~Korea University}\\
~~~~~~~~~Seoul, Republic of Korea \\
~~~~~~~~~dan\_li@korea.ac.kr}
\and
\IEEEauthorblockN{~~Hye-Bin Shin}
\IEEEauthorblockA{\textit{~~~~Dept. of Brain and Cognitive Engineering} \\
\textit{Korea University}\\
Seoul, Republic of Korea \\
hb\_shin@korea.ac.kr}
\linebreakand
\IEEEauthorblockN{Kang Yin}
\IEEEauthorblockA{\textit{Dept. of Artificial Intelligence} \\
\textit{Korea University}\\
Seoul, Republic of Korea \\
charles\_kang@korea.ac.kr}
\and
\IEEEauthorblockN{Seong-Whan Lee}
\IEEEauthorblockA{\textit{Dept. of Artificial Intelligence} \\
\textit{Korea University}\\
Seoul, Republic of Korea \\
sw.lee@korea.ac.kr}
}

\maketitle

\begin{abstract}
The significant inter-subject variability in electroencephalogram (EEG) signals often results in substantial changes to neural network weights as data distributions shift. This variability frequently causes catastrophic forgetting in continual EEG decoding tasks, where previously acquired knowledge is overwritten as new subjects are introduced. While retraining the entire dataset for each new subject can mitigate forgetting, this approach imposes significant computational costs, rendering it impractical for real-world applications. Therefore, an ideal brain-computer interface (BCI) model should incrementally learn new information without requiring complete retraining, thereby reducing computational overhead.
Existing EEG decoding methods typically rely on large, centralized source-domain datasets for pre-training to improve model generalization. However, in practical scenarios, data availability is often constrained by privacy concerns. Furthermore, these methods are susceptible to catastrophic forgetting in continual EEG decoding tasks, significantly limiting their utility in long-term learning scenarios.
To address these issues, we propose the Personalized Continual EEG Decoding (PCED) framework for continual EEG decoding. The framework uses Euclidean Alignment for fast domain adaptation, reducing inter-subject variability. To retain knowledge and prevent forgetting, it includes an exemplar replay mechanism that preserves key information from past tasks. A reservoir sampling-based memory management strategy optimizes exemplar storage to handle memory constraints in long-term learning.
Experiments on the OpenBMI dataset with 54 subjects show that PCED balances knowledge retention and classification performance, providing an efficient solution for real-world BCI applications.

\end{abstract}


\section{INTRODUCTION}
Brain-computer interfaces (BCIs) have gained significant traction in medical rehabilitation, offering innovative solutions for individuals with motor impairments or those recovering from strokes. These systems enable users to control external devices, such as robotic arms, through motor imagery (MI) \cite{prabhakar2020framework,edelman2019noninvasive,han2020classification,mao2019brain,mane2020multi}, or to express intentions via imagined speech without vocalization \cite{garcia2023intra}. Furthermore, electroencephalogram (EEG) signals are increasingly applied in detecting mental states \cite{yu2019weighted,myrden2015effects,jeong2019classification}, such as irregular brain activity related to emotions, facilitating advancements in emotion analysis \cite{ma2022few,kim2015abstract}.

Despite these promising developments, decoding EEG signals remains challenging due to their high variability across individuals and even within the same individual over time, posing significant obstacles to achieving consistent and accurate performance. Techniques such as transfer learning \cite{thinker} and domain adaptation \cite{she2023improved} have been proposed to address these challenges. However, these approaches often rely on large pre-training datasets, which are difficult to obtain in medical contexts due to stringent data privacy concerns. Moreover, these methods \cite{mane2021fbcnet} are prone to catastrophic forgetting (CF), where previously acquired knowledge is overwritten as new data is incorporated during training \cite{french1999catastrophic}.

In an ideal scenario, intelligent systems should acquire new knowledge from sequential data streams while retaining previously learned information—a key aspect of incremental or continual learning, which is fundamental to advancements in artificial intelligence. To address catastrophic forgetting, various strategies have been developed, including regularization-based methods \cite{rosenfeld2018incremental}, network expansion \cite{liu2021adaptive}, and memory replay \cite{xiao2023online}. Among these, memory replay has garnered particular attention for its simplicity and effectiveness.

To enable continual online EEG decoding while addressing the challenges of catastrophic forgetting, we propose a Personalized Continual EEG Decoding Framework (PCED) tailored for continual MI-EEG decoding tasks. The proposed framework leverages Euclidean Alignment to achieve rapid domain adaptation, effectively mitigating the impact of inter-subject variability. To retain prior knowledge and mitigate forgetting, we incorporate an exemplar replay mechanism designed to preserve critical information from previously learned tasks. Furthermore, to address memory constraints inherent in long-term incremental learning, the framework employs a reservoir sampling-based memory management strategy, optimizing the storage and utilization of exemplar data over time. Extensive experimental evaluations on benchmark datasets demonstrate the effectiveness of the proposed PCED in mitigating forgetting and highlight its advantages in continual MI-EEG classification tasks.

\begin{figure}
    \centering
    \includegraphics[width=1\linewidth]{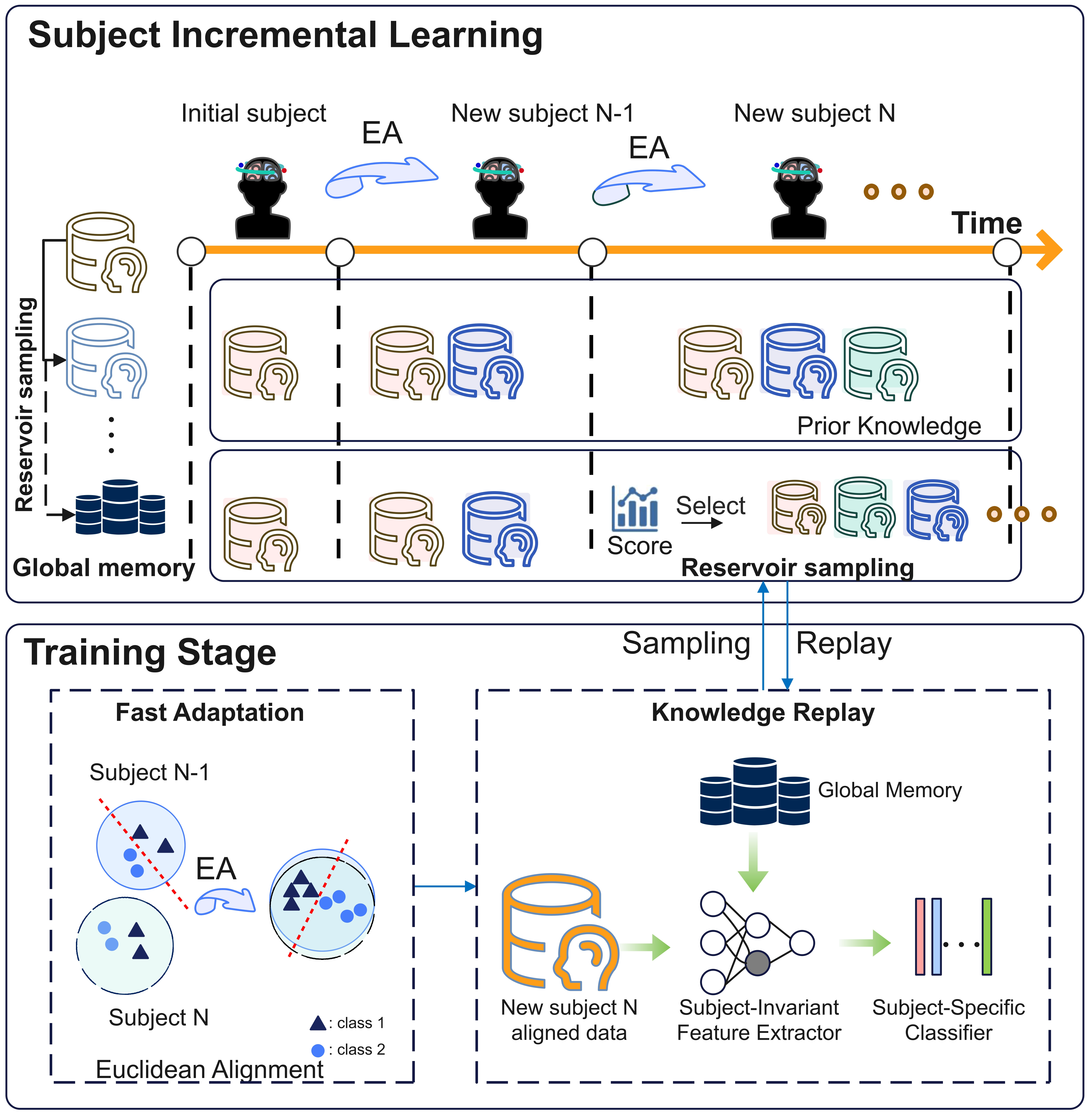}\vspace{-0.2cm}
    \caption{Our personalized incremental learning framework for continual EEG decoding has three key components: (1) EA enables fast adaptation by reducing domain shifts between subjects; (2) memory replay reinforces previous knowledge by replaying stored samples alongside new aligned samples during MI-EEG decoding, preventing forgetting; (3) a reservoir sampling-based memory management strategy maintains a small, dynamic buffer of prior knowledge, using probability scores to efficiently select new samples when memory capacity is exceeded.}
    \label{fig:model framework}
\end{figure}

\section{METHODOLOGY}
\subsection{Problem Definition}
Due to the substantial variability in EEG signals across different subjects, models in continual learning tend to prioritize updating the weights associated with new subjects, leading to the forgetting of previously acquired knowledge. Therefore, our goal is to train a model $\mathcal{F}: \mathcal{X} \rightarrow \mathcal{Y}$ that can continually learn across multiple subjects while minimizing the forgetting of prior knowledge. Formally, let $\mathcal{V} =\left\{\mathcal{D}_1, \mathcal{D}_2, ..., \mathcal{D}_k, ..., \mathcal{D}_N\right\}$  denote the sequential data stream consisting of $N$ subjects. The dataset for the $k^\text{th}$ subject is represented as $\mathcal{D}_k=\left\{(X_k^i, Y_k^i, L_k^i)_{i=1}^{m_k}\right\}$, where $X_k \in \mathcal{X}$ corresponds to the input data, $Y_k \in \mathcal{Y}$ refers to the class label, $L_k$ designates the domain label, and $m_k$ indicates the total number of samples in $\mathcal{D}_k$.

\subsection{Euclidean Alignment for Fast Domain Adaptation}
Euclidean alignment (EA)~\cite{tl-ea2019} is an alignment technique designed to reduce the inter-subject variability of EEG signals by calculating a common reference covariance matrix, thereby improving the robustness of downstream analysis and classification tasks.
Given a subject data $X \in \mathbb{R}^{N \times C \times T}$, where $N$ is the number of trials, $C$ represents the number of channels, and $T$ indicates the number of time points. By normalizing each sample using a common reference covariance matrix, EA improves the consistency of the data, making it more suitable for subsequent analysis and classification tasks. The EA process involves the following steps. First, compute the covariance matrix $R_i$ for each sample $X_i$:
\begin{equation}
R_i = \frac{1}{T-1} \sum_{t=1}^{T} (X_{it} - \mu_i)(X_{it} - \mu_i)^\top,\nonumber 
\end{equation}
where $X_{it}$ is the $t$-th time point of the $i$-th sample, and $\mu_i$ is the mean vector of the $i$-th sample. Next, calculate the reference covariance matrix $\Sigma_{\text{ref}}$ as the mean of the individual covariance matrices, apply the whitening transformation to each sample $X_i$ to obtain the aligned data $X_i^{EA}$: 

\begin{equation}
X_i^{EA} = (\frac{1}{N} \sum_{i=1}^{N} R_i)^{-\frac{1}{2}} X_i.\nonumber 
\end{equation}

\subsection{Exemplar Replay for Prior Knowledge Preservation}
Upon the arrival of the $k^\text{th}$ subject's dataset $\mathcal{D}_k$, we train the subject-specific mapping model $\mathcal{F}_k(X_k) = Y_k$.  After each subject finish training, we randomly store $s$ samples in memory $M$ to mitigate forgetting, which is denoted as  $\left\{x_i\right\}_{i=1}^{s}\sim \left\{\mathcal{D}_j\right\}_{j=1}^{k-1}$. Notably, the model only has access to the data of the current $k^\text{th}$ subject, $\mathcal{D}_k$, and cannot leverage data from previous subjects, $\left\{{\mathcal{D}_1, \mathcal{D}_2, ..., \mathcal{D}_{k-1}}\right\}$. 

\subsubsection{Continual MI-EEG Decoding with Replay Memory}
during the training of the $k^\text{th}$ subject, the stored exemplars from the global memory are combined with the new subject’s data for incremental learning. 
The classification output $y_k$ is optimized using the cross-entropy loss $\mathcal{L}_\text{cls}$, defined as:

\begin{equation} 
\mathcal{L}_\text{cls} = -\mathbb{E}_{(x_k, y_k) \sim (X_k, Y_k) \cup \mathcal{M}_k}\left[\sum_{c=1}^C \mathbbm{1}_{[c=y_k]} \log \sigma(\mathcal{F}_k(x_k))_c\right],\nonumber 
\end{equation}
where $C$ represents the number of MI classes, and $\sigma$ is the softmax function. Through the use of the exemplar replay mechanism, the model effectively mitigates forgetting.




\subsubsection{Memory Management with Reservoir Sampling}
when data from a new subject $k$, denoted as $\mathcal{D}_k$, arrives, the replay mechanism manages the global memory buffer $\mathcal{M}$ using a \textit{reservoir sampling} strategy. The core idea is to randomly retain a fixed number of exemplars from the streaming data, ensuring that the memory buffer does not exceed a predefined capacity. The update process works as follows:

\begin{itemize}
    \item memory update: if the global memory $\mathcal{M}$ has available space (i.e., the buffer size $|\mathcal{M}|$ is less than the capacity $B$), the new samples are directly added to the memory;
    \item reservoir sampling: when the memory is full, the reservoir sampling method \cite{rosenfeld2018incremental} probabilistically replaces existing samples in the memory. The probability of replacing an old sample with a new one is given by:
    \begin{equation}
    p = \frac{B}{B + |\mathcal{M}|},\nonumber 
    \end{equation}
    ensuring that each sample has an equal chance of being retained in memory.
\end{itemize}
Based on the two critical points, we define the update process as:
\begin{equation}
\begin{aligned}
\mathcal{M} &\leftarrow \mathcal{M} \cup \{(x_i, y_i, t_i)\} \quad \text{if} \ |\mathcal{M}| < B, \nonumber \\
\mathcal{M}_j &\leftarrow (x_i, y_i, t_i) \quad \text{with prob. } p = \frac{B}{B + |\mathcal{M}|},\nonumber 
\end{aligned}
\end{equation}   
where $j$ is a uniformly randomly selected index from the memory $\mathcal{M}$, and $x_i$, $y_i$ and $t_i$ are input data, corresponding label and timestamp, respectively.

\section{EXPERIMENTS}\label{sec:experiments}
\subsection{Dataset}
In this study, we evaluated our method using the OpenBMI dataset of 54 subjects \cite{lee2019eeg} for continual subject-incremental MI-EEG classification. EEG signals were recorded from 62 channels at 1,000 Hz, with 20 electrodes selected for motor imagery tasks, following prior research \cite{gitgan,kobler2022spd,bangNNLS-sst}. The data was down-sampled to 250 Hz, and 4-second segments of motor imagery grasping tasks were extracted. We used 70\% of each subject’s data for training, with the remaining 30\% split equally for validation and testing.

\subsection{Evaluation Metrics and Baselines}
We measure a widely used metric backward transfer (BWT) \cite{lopez2017gradient} to assess the effect of new subject learning on previously seen subjects. BWT is calculated as: $\text{BWT} = \frac{1}{N-1} \sum_{i=1}^{N-1} \left( a_{N,i} - a_{i,i} \right)$, where \( a_{j,i} \) is the accuracy on subject \( i \) after training on subject \( j \). Negative BWT indicates forgetting, while positive BWT shows performance improvement on earlier subjects.
We also calculate the average accuracy (ACC) across all subjects after the final round of learning to assess overall retention. 
For the baseline methods, we carefully selected several well-established approaches to provide a comprehensive comparison for our proposed method. Subject-specific finetuning (SFT) refers to the process of incrementally training a model without employing any methods to mitigate forgetting. In this approach, the model's parameters for the new task are initialized based on the parameter values learned from the previous task. Regularization-based methods, such as elastic weight consolidation (EWC) \cite{kirkpatrick2017overcoming}, aim to address this issue by selectively preserving important weights. Additionally, experience replay (ER) \cite{chaudhry2019tiny} is a rehearsal strategy that stores a limited number of replay samples in a memory buffer. 

\subsection{Experimental Setting}
For feature extraction, we utilize four different base models as feature extractors: ShallowConvNet (SCN) \cite{pan2016shallow}, DeepConvNet (DCN) \cite{deepconvnet}, EEGNet-8,2 \cite{EEGnet}, and EEGNet-64,16 \cite{EEGnet}, where the numbers denote the kernel sizes. Simultaneously, we employ a three-layer Multi-Layer Perceptron as the domain discriminator, incorporating softmax and ELU activation functions to predict both the subject index and MI class labels. In our problem setting, data from unseen subjects is not available in advance. Note that, we do not use any pre-trained models for feature extraction. Instead, we use a class-balanced memory buffer, storing 10 samples per class, to reinforce knowledge during training. We apply a learning rate of 0.001 to train the entire network over 200 epochs, and to prevent overfitting, we employ an early stopping mechanism.

\begin{table}[t]
\caption{Comparative study on the performance of different models and backbones for the OpenBMI dataset.}
\label{table:PCED}
\resizebox{\linewidth}{!}{%
\begin{tabular}{llll}
\toprule
Method & Backbone & ACC (\%) & BWT (\%) \\
\midrule
\multirow{4}{*}{SFT}

  & SCN \cite{deepconvnet}   & 50.37 ± 0.33 &-14.93 ± 1.14 \\ 
  & DCN \cite{deepconvnet} & 50.76 ± 0.44 &-20.33 ± 0.26 \\
    & EEGNet-8,2\cite{EEGnet}   &  49.81 ± 0.67 &-17.04 ± 0.18 \\
  & EEGNet-64,16\cite{EEGnet}   & 49.70 ± 0.69 &-22.50 ± 0.29 \\ 
\midrule
\multirow{4}{*}{EWC\cite{kirkpatrick2017overcoming}}

  & SCN \cite{deepconvnet}   &  66.31 ± 0.53 &-4.01 ± 0.34  \\ 
  &  DCN \cite{deepconvnet} &  67.93 ± 0.13 &-4.67 ± 0.90  \\
    & EEGNet-8,2\cite{EEGnet}   &  65.93 ± 0.93 &-5.17 ± 0.89      \\
  & EEGNet-64,16\cite{EEGnet}   &  66.10 ± 0.21 &-5.01 ± 0.93    \\ 
\midrule
\multirow{4}{*}{ER\cite{chaudhry2019tiny}}

  & SCN \cite{deepconvnet}   & 74.46 ± 0.91 &-5.58 ± 0.67 \\ 
  &  DCN \cite{deepconvnet} &74.20 ± 0.26 &-2.98 ± 0.58\\ 
    & EEGNet-8,2\cite{EEGnet}   & 74.49 ± 0.33 &-1.49 ± 0.60 \\
  & EEGNet-64,16\cite{EEGnet}   & 74.11 ± 0.38 &-2.24 ± 0.37\\ 
\midrule
\multirow{4}{*}{PCED (Ours)}

  & SCN \cite{deepconvnet}& 79.89 ± 0.63 &-1.85 ± 1.03 \\
  &  DCN \cite{deepconvnet} & \textbf{82.27 ± 0.43} & \textbf{~0.22 ± 0.48 } \\
    & EEGNet-8,2\cite{EEGnet} & 79.89 ± 0.59 &-2.99 ± 0.56  \\
  & EEGNet-64,16\cite{EEGnet} & 79.93 ± 0.23 &-3.17 ± 0.38 \\
\bottomrule
\vspace{0.1cm}
\end{tabular}
}
\footnotesize{{$^*$ACC: average accuracy, BWT:  backward transfer, SFT: subject-specific finetuning.}}
\end{table}

\subsection{Results and Discussion}
Our results show that in the subject-incremental continual MI-EEG classification task, our proposed method consistently outperforms other approaches. As illustrated in Table \ref{table:PCED}, regardless of the backbone architecture used, the proposed PCED method achieved the highest classification accuracy and lowest forgetting rate. Notably, when paired with DCN, PCED yielded higher ACC and BWT than with other backbones, suggesting that improving feature extraction can further boost the learning performance of our method.

\begin{figure}
    \centering
        \resizebox{\linewidth}{!}{

    \includegraphics[width=1\linewidth]{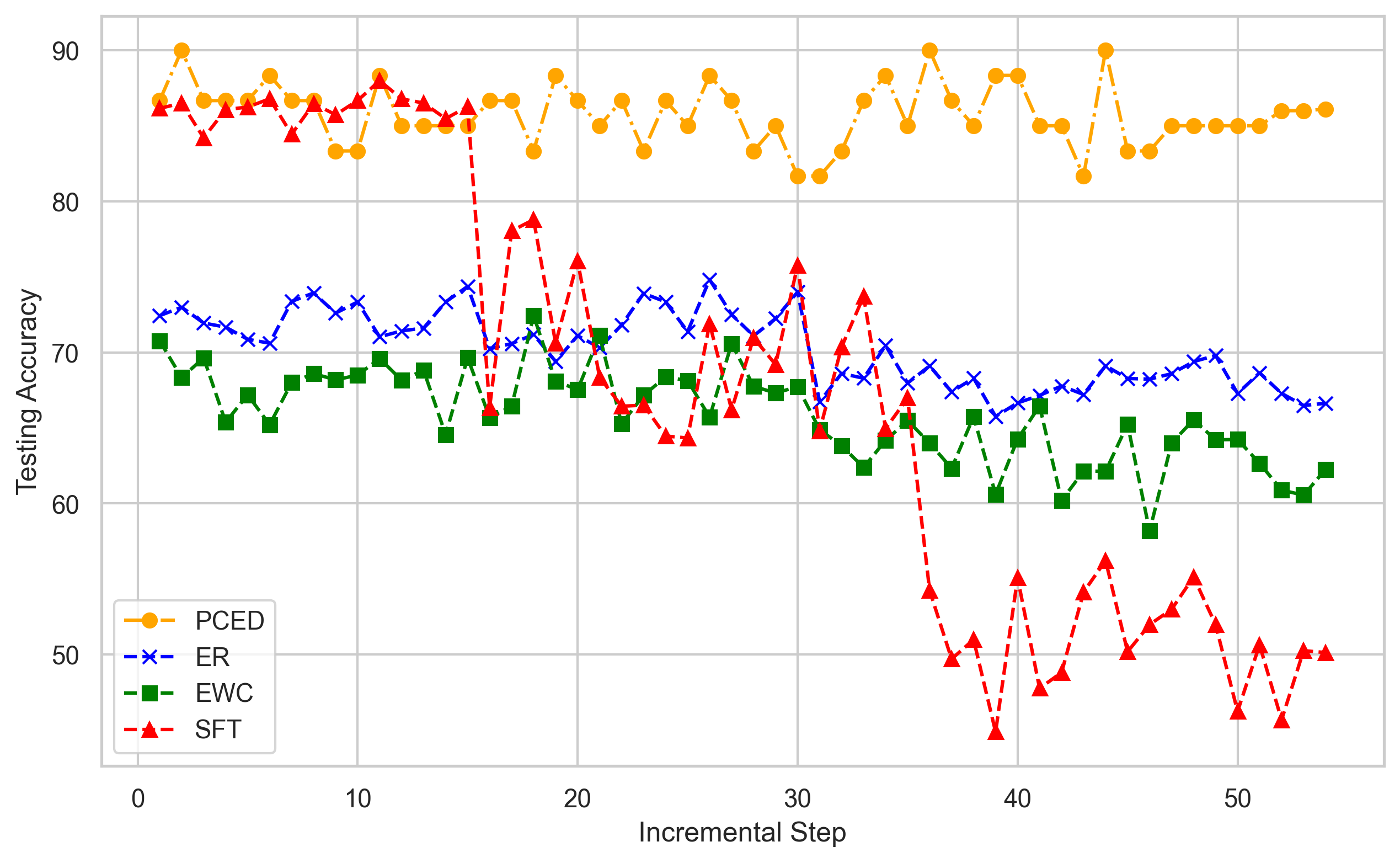}}
    \caption{Testing accuracy of different methods at each incremental learning stage for the initial subject in continual EEG decoding.}
    \label{fig:fogetting curve}
\end{figure}

To further illustrate the effectiveness of PCED in preventing forgetting during incremental learning, we plotted the forgetting curve in Fig. \ref{fig:fogetting curve}, which tracks test accuracy for a representative subject after incremental learning across 54 subjects. As shown, our method maintains a smoother forgetting curve compared to the finetuning approach, which lacks strategies to prevent forgetting, highlighting the risk of CF without continual learning techniques.

\section{CONCLUSION}
In this paper, we present the Personalized Continual EEG Decoding (PCED) framework, designed for continual MI-EEG classification tasks. The framework employs Euclidean Alignment to address domain shifts between subjects, enabling smoother knowledge transfer as new subjects are introduced. To mitigate catastrophic forgetting, we integrate a memory replay mechanism that retains previously acquired knowledge while providing flexible control over the number of replayed samples, significantly reducing memory overhead—critical for long-term learning with an increasing number of subjects. Experimental results demonstrate that PCED outperforms baseline methods by effectively mitigating memory loss and maintaining high classification performance in subject-incremental learning scenarios. Additionally, the framework addresses key challenges in continual learning for MI-EEG data, including memory limitations and domain variability. Overall, PCED offers a robust and scalable solution for continual MI-EEG decoding, preserving knowledge from prior subjects while efficiently adapting to new ones, advancing personalized and flexible brain-computer interface (BCI) applications.


\bibliographystyle{IEEEtran}


\bibliography{REFERENCE}


\end{document}